\input harvmac

\def\half{{1\over 2}}

\def\sh{\hat{\sigma}}

\def\la{{\Lambda}}



\Title{}{\vbox{\centerline{Cardy-Verlinde Formula and Holographic Dark Energy}}}

\centerline{Ke Ke$^1$ and  Miao Li$^{1,2}$ } \vskip .5cm
\centerline{\it $^1$ Institute of
Theoretical Physics} \centerline{\it Academia Sinica, P.O. Box
2735}
\centerline{\it Beijing 100080, China} \centerline{\it and}
\centerline{\it $^2$ Interdisciplinary Center of Theoretical Studies}
\centerline{\it Academia Sinica, Beijing 100080, China}
\centerline{\tt kek@itp.ac.cn, mli@itp.ac.cn}

\bigskip

If we separate energy in a holographic theory into an extensive part
and an intrinsic part, where the extensive part is given by the cosmological
constant, and assume entropy be given by the Gibbon-Hawking formula, the
Cardy-Verlinde formula then implies an intrinsic part which agrees with
a term recently proposed by Hsu and Zee. Moreover, the cosmological constant
so derived is in the form of the holographic dark energy, and the coefficient
is just the one proposed recently by Li. If we replace the entropy by the
so-called Hubble bound, we show that the Cardy-Verlinde formula is the same
as the Friedmann equation in which the intrinsic energy is always dark energy. We
work in an arbitrary dimension.

\Date{July, 2004}



\nref\snc{A. G. Riess et al., Astron. J. 116 (1998) 1009; S.
Perlmutter et al., APJ 517 (1999) 565.}
\nref\ben{C. L. Bennett et
al., astro-ph/0302207, Astrophys.J.Suppl. 148 (2003) 1; D. N. Spergel et al., astro-ph/0302209, Astrophys.J.Suppl. 148 (2003) 175; H.
V. P. Peiris et al., astro-ph/0302225, Astrophys.J.Suppl. 148 (2003) 213.} \nref\ckn{A. Cohen,
D.~Kaplan and A.~Nelson, hep-th/9803132, Phys. Rev.  Lett. 82
(1999) 4971.} \nref\st{S. Thomas, Phys. Rev. Lett.  89 (2002)
081301.}
\nref\sh{S. D. H. Hsu, Phys.Lett. B594 (2004) 13-16.}
\nref\mli{M. Li, hep-th/0403127, Phys. Lett. B603 (2004) 1.}
\nref\hdark{ Q. G. Huang and Y. Gong, astro-ph/0403590, JCAP 0408 (2004) 006; Y. Gong, hep-th/0404030, Phys. Rev. D 70 (2004) 064029;
Q. G. Huang and M. Li, astro-ph/0404229, JCAP 0408 (2004) 013;
M. Ito, hep-th/0405281;  K. Enqvist and M. S. Sloth, hep-th/0406019.}
\nref\hz{S. Hsu and A. Zee,
hep-th/0406142.}
\nref\verl{E. Verlinde, hep-th/0008140.}
\nref\cai{R. G. Cai, hep-th/0111093, Phys. Lett. B525 (2002) 331.}
\nref\rev{B. Wang, E. Abdalla, R.-K. Su, hep-th/0101073,
Phys.Lett. B503 (2001) 394;
S Nojiri, S D Odintsov and S Ogushi, hep-th/0205187, Int.
Journ. Mod. Phys. A17 (2002) 4809.}

Recently there has been considerable interest in explaining the observed
dark energy \refs{\snc,\ben} by the holographic dark energy model \ckn. (See
also \st.)  A. Cohen and collaborators suggested sometime ago that
the zero-point energy in quantum field theory is affected by an
infrared cut-off $L$, thus, the cosmological constant is given by
a formula $\Lambda \sim M_p^2L^{-2}$. If one takes $L^{-1}$ be the
Hubble constant $H$, then dark energy is close to the critical
energy density observed. However, Hsu pointed out that in a
universe with this dark energy and matter, this formula yields a
wrong equation of state \sh. Li subsequently proposed that the
infrared cut-off $L$ should be given by the size of event horizon
\mli , namely $L=a\int_t^\infty dt'/a(t')$, then the observed data
can be nicely fitted. This model was further studied in
\refs{\hdark}

Hsu and Zee recently made the following interesting observation \hz:
Suppose there be an infrared cut-off $L$ in an effective action
describing our universe, the cosmological constant term
contributes an term $\la L^4$ to this action, suppose further for some reason
(maybe due to quantum corrections, as suggested by these authors)
there be an additional term $M_p^4/\la$ in this effective action,
then minimizing the action yields $\la=M_p^2L^{-2}$. Hsu and Zee
propose the following picture, let the energy scale associated with
the cosmological constant be $m_\la$, the energy scale associated with
the infrared cut-off be $m_U$, thus $m_\la=\sqrt{M_pm_U}$, a formula
reminiscent of the seesaw mechanism.

In Cohen's paper \ckn, the holographic form of dark energy is
argued for by setting the UV and IR cutoff to saturate the bound
set by formation of a black hole. In this note, we assume that the
dark energy has a holographic origin. By relating it to the
Cardy-Verlinde formula, we thus argue for he holographic form of
dark energy from another viewpoint.

Verlinde suggested, motivated both by holography and by Cardy's formula for a
2 dimensional conformal field theory, a generalization
of Cardy's formula to a $n$ dimensional conformal field theory \verl
\eqn\cvf{S={4\pi R\over n-1}\sqrt{E_cE_e},}
where $S$ is entropy, $E_c$ is a part of energy similar to Casimir
energy, and $E_e$ is the extensive part of energy (our convention
for $E_c$ differs from that of Verlinde by a factor of $\half$). The
CFT is supposed to live in spacetime with topology $R\times S^{n-1}$,
$n-1$ is the spatial dimension.

The effective action proposed in \hz\ can be written in a form of
effective energy \eqn\hze{E=\la L^3+{M_p^4\over \la L}.}
Minimizing the effective energy also results in the correct
scaling for $\la$. $\la L^3$ is of course the extensive energy in
bulk, it is reasonable to assume that it is also the extensive
energy in the field theory dual to the cosmology theory in
question, since by dimensional analysis, $\rho=\la L$ can be
viewed as the $2d$ energy density, then $E_e=\rho L^2$ scales
correctly in terms of the $2d$ volume. $M_p^4/(\la L)$ is the
intrinsic (Casimir) energy, then $\sqrt{E_cE_e}L=M_p^2L^2$, a
quantity resembling the Gibbons-Hawking entropy if we identify $L$
with $R$, the size of the cosmic horizon. This remarkable
coincidence strongly suggests to us that the new term $M_p^4/(\la
L)$ has a holographic origin: it arises as the Casimir energy in a
dual theory, if so, its nature is indeed quantum mechanical.

Note that while Verlinde originally works with a CFT with
spacetime dimension the same as that of the whole universe, in our
problem we prefer to interpret his formula as the one in a dual
theory with one fewer dimension. There are many papers applying
the Cardy-Verlinde formula to dS space, in the spirit of dS/CFT
correspondence. For example, in \cai, the author suggested that
there is a CFT theory dual to dS and checked the Cardy-Verlinde
formula. Here, we also assume the dark energy in the bulk dual to
a CFT theory on the boundary. There is a correspondence between
the extensive holographic dark energy in the bulk and extensive
energy in the CFT theory. So we can use the Cardy-Verlinde formula
\cvf. Of course, our purpose as well as the definition of energy
are completely different from those in \cai. Other discussions on
relation between the Cardy-Verlinde formula and the Friedmann
equation with a cosmological constant can be found for instance in
\rev.

In Li's paper \mli, there is an undetermined parameter $c$. The
author suggested $c=1$ through an argument compared to blackhole.
Here, we will use the Cardy-Verlinde formula to determine this
parameter. Let us work instead in an arbitrary dimension and be
more accurate numerically. Suppose in our universe there be a
cosmic horizon of size $R$, or more generally, an infrared cut-off
$R$. Let spacetime be $n+1$ dimensional. The Gibbons-Hawking
entropy is \eqn\ghe{S_{GH}={\Omega_{n-1}R^{n-1}\over
4G}=2\pi\Omega_{n-1} M_p^{n-1}R^{n-1},} where $\Omega_{n-1}$ is
the volume of the unit sphere $\Omega^{n-1}$, $M_p^{n-1}=1/(8\pi
G)$ the reduced Planck mass in $n+1$ dimension. Note that we need
to use the Gibbons-Hawking entropy for our purpose, this is why we
interpret $R$ as the size of the cosmic horizon.

Next, with the presence of dark energy density $\la$, we propose
that the bulk energy be $E_e=\la V$, where $V={\Omega_{n-1}\over n}R^n$ is
the volume enclosed by the cosmic horizon, and
interpret this as the extensive energy in a dual theory. It is
certainly difficult to regard $E_e$ as energy in the flat patch
$ds^2=-dt^2+a(t)^2(dx^i)^2$, since there is no notion of conserved
energy in this system of coordinates. If we work in the static
coordinates
\eqn\smet{ds^2=-(1-{r^2\over R^2})dt^2+(1-{r^2\over R^2})^{-1}dr^2
+r^2d\Omega_{n-1}^2,}
then $E=\int \sqrt{G_{00}G_{rr}}\la=\la V$ is the conserved energy
conjugate to $t$, the proper time of the comoving observer sitting
at $r=0$. However, as we shall explain later, it still makes sense to interpret
$E$ calculated in the flat patch as energy in the dual theory.

Let $E_c$ be unknown, applying the Cardy-Verlinde formula \cvf\ to
$E_e$, $S_{GH}$ and $E_c$ we solve $E_c$ in terms of $E_e$ and
$S_{GH}$ and find \eqn\cef{E_c={n(n-1)^2\over
4\la}\Omega_{n-1}M_p^{2n-2}R^{n-4}.} This is a $n+1$ dimensional
generalization of proposal in \hz. Note that we assume that the
Cardy-Verlinde formula is used in a $n$ dimensional dual theory.
In the dual theory, the total energy is then
\eqn\te{E=E_e+E_c={1\over n}\Omega_{n-1}\la R^n+{n(n-1)^2\over
4\la}\Omega_{n-1} M_p^{2n-2}R^{n-4}.} Minimizing $E$, we obtain
\eqn\lae{\la ={n(n-1)\over 2}M_p^{n-1}R^{-2},} the holographic
dark energy in $n+1$ dimension. Take $n=3$, this result agrees
with the one proposed in \mli\ with parameter $c=1$ advocated
there.

It is easy to check that the above result is consistent with a $n+1$
dimensional de Sitter space. The Friedmann equation in $n+1$
dimensions is
\eqn\fe{{n(n-1)\over 2}M_p^{n-1}H^2=\rho.}
Replacing $\rho$ by $\la$ and plugging \lae\ into the above equation,
we have $H^2=R^{-2}$, this is exactly the correct relation between the Hubble
constant $H$ and the cosmic horizon size $R$ for a de
Sitter space.

Having succeeded in deriving the dark energy formula in a holographic dual,
we now come to a universe with both dark
energy and other form of energy. If we insist on the Cardy-Verlinde
formula, since at least the extensive energy is larger than the
one corresponding to dark energy, entropy should be greater than
the Gibbons-Hawking entropy. This entropy, just like the one in
a closed universe discussed in \verl, must be the Hubble bound $S_H$.
We propose the following formula for the Hubble bound
\eqn\hef{S_H={nHV\over 4G}=2\pi\Omega_{n-1}M_p^{n-1}HR^n,}
where we work in the flat patch, only in this case $H$ is defined.
This formula is similar to the one proposed in \verl, but with
a slight difference in coefficient since we have a flat universe
with boundary. We may or may not interpret $R$ as the size of event
horizon, but it must be the same infrared cut-off appearing in the
dark energy formula. In a pure de Sitter space, the Hubble bound agrees
with the Gibbons-Hawking entropy only when $R$ is taken to be the
horizon size.

Solving $H$ in terms of $S_H$ \eqn\hinv{H^2={S_H^2\over (2\pi
\Omega_{n-1}M_p^{n-1}R^n)^2}.} On the L.H.S. we use the Friedmann
equation \eqn\fffr{M_p^{n-1}H^2={2\over n(n-1)}(\rho_\la+\rho_m),}
where $\rho_\la$ is the holographic dark energy \lae, and $\rho_m$
is whatever energy density in question, we deduce
\eqn\dee{\eqalign{S_H^2&={8\pi^2M_p^{n-1}R^{2n}\Omega_{n-1}^2\over
n(n-1)} (\rho_\la +\rho_m) \cr &=({4\pi R\over
n-1})^2E_c(E_c+E_m),}} where $E_c=\la V=E_e$ is the ``on-shell"
Casimir energy associated with $\la$, and $E_c+E_m=E_e+\rho_m V$
is the total extensive energy. We thus find that the
Cardy-Verlinde formula is equivalent to Friedmann equation
provided we always use the same formula for the Casimir energy
without any contribution from the matter part. This result is
similar to Verlinde's: When the Casimir energy is replaced by
Bekenstein-Hawking energy(in our case, Bekenstein-Hawking energy
is similar to the on-shell Casimir energy), the Cardy-Verlinde
formula is equals to the Friedman equation.

Note that the presumed holographic dual itself hides the time evolution
in the bulk, the Friedmann equation governs time evolution in the
bulk, but its interpretation in the dual theory is a static relation.
Similarly, one can not ask the question as to why the total energy is not
conserved in the bulk. By the same token, $S_H$ is interpreted in the
dual theory as total entropy, while in the bulk it is interpreted as
a upper bound only. One can not take $S_H$ as the real entropy in the
bulk, since in general $S_H$ viewed as a function of time does not observe
the second law of thermodynamics. To see this, let us compute
\eqn\secl{{d\over dt}(HR^n)=R^{n-2}\left(-1+\sqrt{\Omega_\la}-{n\over 2}
(1+w){1-\Omega_\la\over\Omega_\la}+n({1\over\Omega_\la}-{1\over\sqrt{\Omega_\la}})
\right),}
where $\Omega_\la$ is the fraction of the dark energy, $w$ is the equation of state
index of matter. One can check numerically that in later time when $\Omega_\la$
approaches $1$, the above quantity can be negative.

Our understanding of this issue is the following. Just as the total energy $E_e
+E_m$, $S_H$ can not be interpreted as a physical quantity in the bulk, its role
is just a bound on entropy. We propose that the Gibbons-Hawking entropy is real
entropy in the bulk, and it is certainly bounded by the Hubble bound. The
Gibbons-Hawking entropy always increases with time.

To summarize, we used the Cardy-Verlinde formula twice. In the first usage,
we apply the Gibbons-Hawking entropy and obtain the Casimir energy, by
minimizing the total energy we obtain the dark energy formula. In the
second usage, we apply the Hubble entropy and obtain the Friedmann equation.
The first usage is valid both for an ``empty" de Sitter space as well
as a spacetime with other form of energy, if this application of
the Cardy-Verlinde formula is correct, we effectively ``derived" the
formula proposed in \mli. The second usage can be regarded as a definition
of the Hubble entropy, thus less conjectural and less consequential.

Acknowledgments.

We are grateful to R. G. Cai and A. Zee for useful
discussions. This work was supported by a grant of
CAS and a grant of NSFC.

\listrefs
\end